\documentclass[lettersize,journal]{IEEEtran}
\usepackage{amsmath,amsfonts}
\usepackage{algorithmic}
\usepackage{algorithm}
\usepackage{array}
\usepackage[caption=false,font=normalsize,labelfont=sf,textfont=sf]{subfig}
\usepackage{textcomp}
\usepackage{stfloats}
\usepackage{url}
\usepackage{verbatim}
\usepackage{graphicx}
\usepackage{cite}
\usepackage{xcolor}

\begin{document}

\title{Readout of Microwave Kinetic Inductance Detector (MKID) Arrays for Habitable Worlds Observatory Using a Polyphase Filterbank Algorithm}

\author{Oketa~Basha$^{*1,3}$, 
        Tracee~Lynn~Jamison-Hooks$^1$,
        Philip~Mauskopf$^1$,
        Lynn~Miles$^2$,
        Sanetra~Newman-Bailey$^2$,
        Abarna~Karthikeyan$^1$,
        Mohammad~Samad$^3$,
        Mariya~Taylor$^3$,
        Sarah~E.~Kay$^4$,
        Sean~Bryan$^1$,
        Devika~Band$^5$, 
        Thomas~Essinger-Hileman$^2$,
        Sumit~Dahal$^2$, 
        Adrian~Sinclair$^2$, 
        Caleb~Distel$^5$,

\thanks{$^1$Arizona State University (ASU), School of Earth and Space Exploration (SESE), Tempe, AZ, USA}
\thanks{$^2$NASA Goddard Space Flight Center (GSFC), Instrument Electronics Development Branch, Greenbelt, MD, USA}
\thanks{$^3$ASU Ira A. Fulton Schools of Engineering (FSE), Tempe, AZ, USA}
\thanks{$^4$Science Systems \& Applications Inc., Lanham, MD, USA}
\thanks{$^5$ASU School of Computing and Augmented Intelligence (SCAI), Tempe, AZ, USA}
\thanks{Corresponding author: tljamiso@asu.edu}
\thanks{Manuscript received September 24th, 2025; revised MMMM DD, 2025.}
\thanks{NASA Strategic Astrophysics Technology (SAT) 2025 Grant 80NSSC25K7184 }}

\markboth{IEEE Transactions on Applied Superconductivity,~Vol.~NN, No.~NN, MMMM~2025}%
{Shell \MakeLowercase{\textit{et al.}}: A Sample Article Using IEEEtran.cls for IEEE Journals}

%\IEEEpubid{0000--0000/00\$00.00~\copyright~2025 IEEE}

\maketitle
\begin{abstract}
The \textit{Habitable Worlds Observatory (HWO)}, a next-generation ultraviolet/optical/infrared space telescope, will require detector technologies capable of supporting substantially larger pixel-count arrays than those flown on previous missions. Microwave Kinetic Inductance Detectors (MKIDs) provide a scalable solution through microwave multiplexing and have already been demonstrated in balloon-borne instruments using Field-Programmable Gate Arrays (FPGAs) for real-time signal processing. A central element of MKID readout is the Polyphase Filter Bank (PFB) spectrometer, which converts digitized time-domain signals into finely resolved frequency channels for subsequent analysis. To meet the demands for broader bandwidths and higher spectral resolution driven by emerging science goals, efficient FPGA-based implementations of the PFB are essential. This work presents current results from a fixed-point digital design methodology for deploying the PFB architecture on space-qualified FPGAs. The approach emphasizes efficient resource utilization and numerical precision while satisfying stringent performance constraints, enabling scalable, high-resolution spectral processing for future space observatories and remote sensing applications.
\end{abstract}


\section{Introduction}
\IEEEPARstart NASA’s next flagship mission, the Habitable Worlds Observatory (HWO; \cite{nasaHWO}), is designed to spectrally characterize Earth-like exoplanets, one of the most demanding science goals of the coming decades. Achieving this objective requires observations in the ultraviolet, optical, and near-infrared (UVOIR) waveband and detectors with extremely low dark count rates. Transition-edge-sensed (TES; \cite{introTES}) bolometers and microwave kinetic inductance detectors (MKIDs;\cite{day2003broadband}) are naturally excellent photon-counting technologies, with MKIDs in particular demonstrating dark count rates on the order of $10^{-3}$ counts/sec/pixel \cite{hwoexoplanets}. 

\begin{figure}[h!]
\centering
\includegraphics[width=9cm]{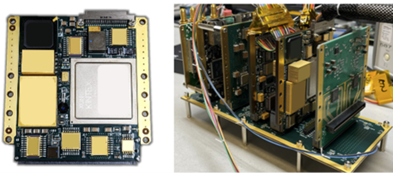}
\caption{The proposed readout system builds upon high-TRL modular SpaceCube hardware. Left: SpaceCube v3.0 Mini FPGA card (primary side pictured) with a Xilinx Kintex KU060. Right: full SpaceCube electronics system with 1U CubeSat cards inserted into a backplane. \cite{casper_toolflow}, \cite{jamisonhooks2025developmentspacequalifiedsignal} }
\label{figure:SpaceCube}
\end{figure}

Our signal-processing readout methodology is designed to monitor the frequency and phase response of large MKID arrays in real time. At their core is a digital signal processing (DSP) chain that performs the essential readout functions, including waveform generation, channelization, cosmic ray removal, tone-tracking, pulse detection, averaging, and data transfer \cite{jamisonhooks2025developmentspacequalifiedsignal} via SpaceWire \cite{parkes2005spacewire}. This functionality is implemented on proven SpaceCube hardware ($\geq$ TRL 6), the same architecture adapted for the PRobe far-Infrared Mission for Astrophysics (PRIMA; \cite{prima}), which provides the baseline for extending the readout design to the Habitable Worlds Observatory. SpaceCube, illustrated in Figure~\ref{figure:SpaceCube}, is a NASA Goddard-developed processor family that established a hybrid approach combining commercial and radiation-hardened technologies \cite{nasaSpaceCube}.

\begin{figure*}
\begin{center}
\begin{tabular}{cc}
\includegraphics[height=0.4\textwidth, clip=true, trim=0in 0in 0in 0in]{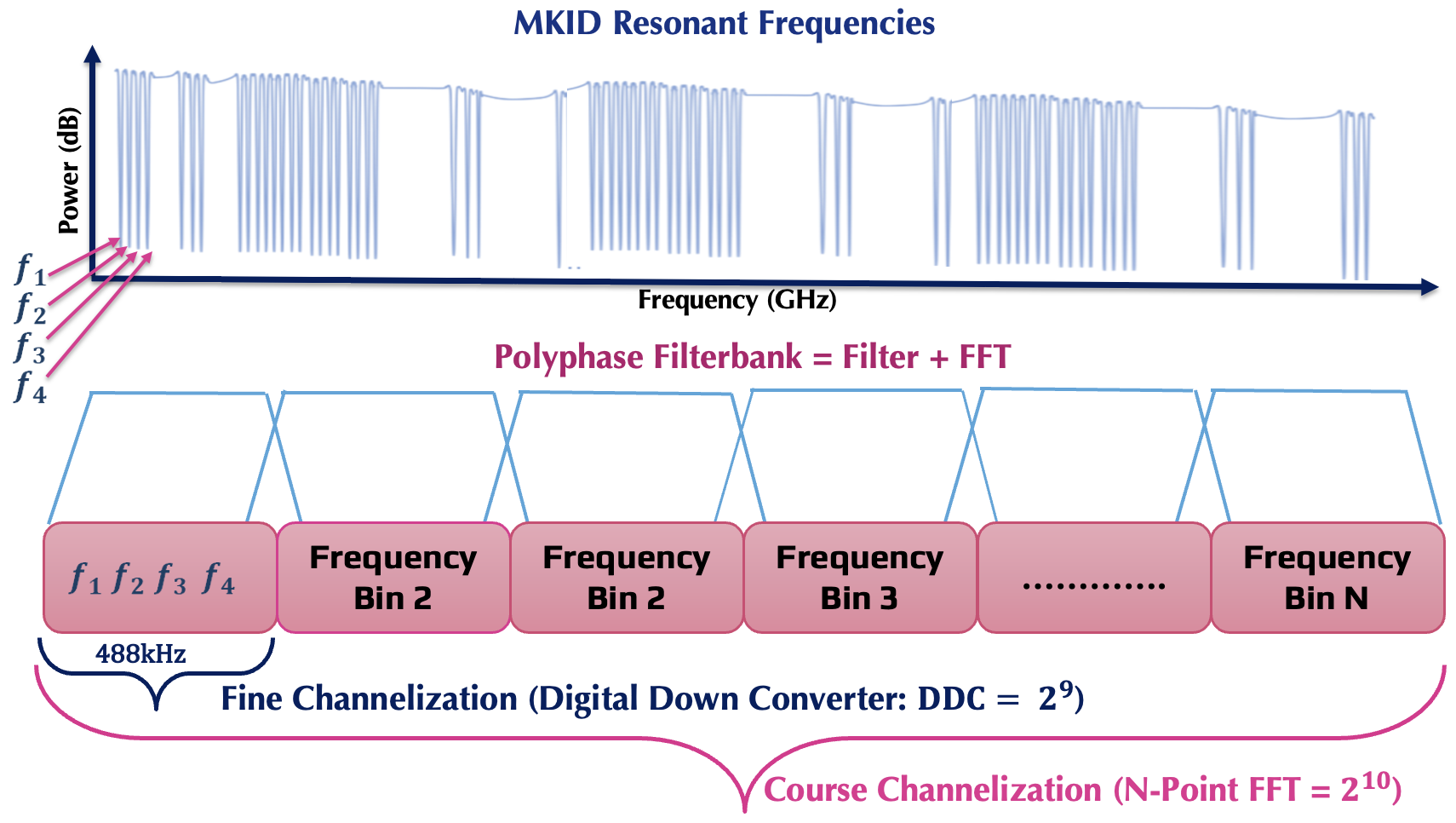} &
\end{tabular}
\caption{A two-stage frequency translation algorithm (coarse and fine channelization) is used to efficiently achieve the required 10~kHz spectral resolution for MKID readout. \cite{casper_toolflow},
\cite{jamisonhooks2025developmentspacequalifiedsignal}}
\label{figure:MKID_PFB}
\end{center}
\end{figure*}

Here, we present recent developments in the primary frequency translation stage, \textit{coarse channelization}, implemented as a polyphase filter bank (PFB)\cite{jamisonhooks2025developmentspacequalifiedsignal}. The PFB builds upon the FPGA-based implementation originally developed for the \textit{Soil Moisture Active Passive (SMAP)} mission \cite{piepmeier2013radio}.

The PFB architecture combines a prototype low-pass filter, decomposed into its polyphase components, with a 1024-point Fast Fourier Transform (FFT). This configuration shapes each frequency bin's response to suppress spectral leakage, a critical requirement for accurate MKID readout. By improving frequency resolution and channel isolation, the PFB enables stable and efficient spectral analysis across densely packed MKID arrays. As such, it plays a central role in advancing space-qualified readout systems for large-format MKID detectors \cite{bradley2021advancements}.

\section{Motivation for a Polyphase Filter Bank Architecture in MKID Readout}

The requirements of the MKID detector ultimately drive the design of the FPGA-based science measurement algorithm. To cover a science bandwidth of \( 2.4\,\text{GHz} \), we require a sampling frequency of \( F_s = 5\,\text{GHz} \). This satisfies the Nyquist criterion and ensures that the full signal range is accurately captured. Due to the uniformity of the MKID architecture and signal characteristics, defining the precision requirements for measuring a single resonator effectively establishes the requirements for all resonators in the MKID array. A single resonator has a bandwidth of 30 kHz. The tone precision placement requirement, which specifies the frequency resolution needed to accurately detect each distinct resonator tone, is set to one-third of the resonator bandwidth, or 
\[
\Delta f = 10\,\text{kHz}.
\]

This resolution requirement ($\Delta f = 10 \, \text{kHz}$) and the Sampling Frequency ($F_{s}$=5\,GHz ) inform the necessary observation window length, or equivalently the number of samples required for tone recovery via FFT-based analysis:

\begin{equation}
    N = \frac{F_s}{\Delta f} = \frac{5 \times 10^9}{10 \times 10^3} = 2^{19} \, \text{samples}.
\end{equation}

Hence, each science measurement must process at least \(2^{19}\) samples to achieve the desired frequency discrimination. This defines the baseline frequency resolution and temporal window required for MKID tone detection and spectral analysis across the full MKID array \cite{jamisonhooks2025developmentspacequalifiedsignal}.
\medskip

The frequency content of MKID detectors is fundamental to accurate signal measurement. Achieving the required frequency resolution would necessitate a direct FFT of length \(2^{19}\). However, such an implementation is impractical for space-qualified FPGAs due to its prohibitive resource requirements. Moreover, a large FFT is inefficient in this application, as many frequency bins would be unused because no MKID resonators fall within them.

To address this, the HWO readout architecture employs a two-stage frequency analysis approach, as illustrated in Figure~\ref{figure:MKID_PFB}.

\begin{figure}[h!]
\centering
\includegraphics[width=9cm]{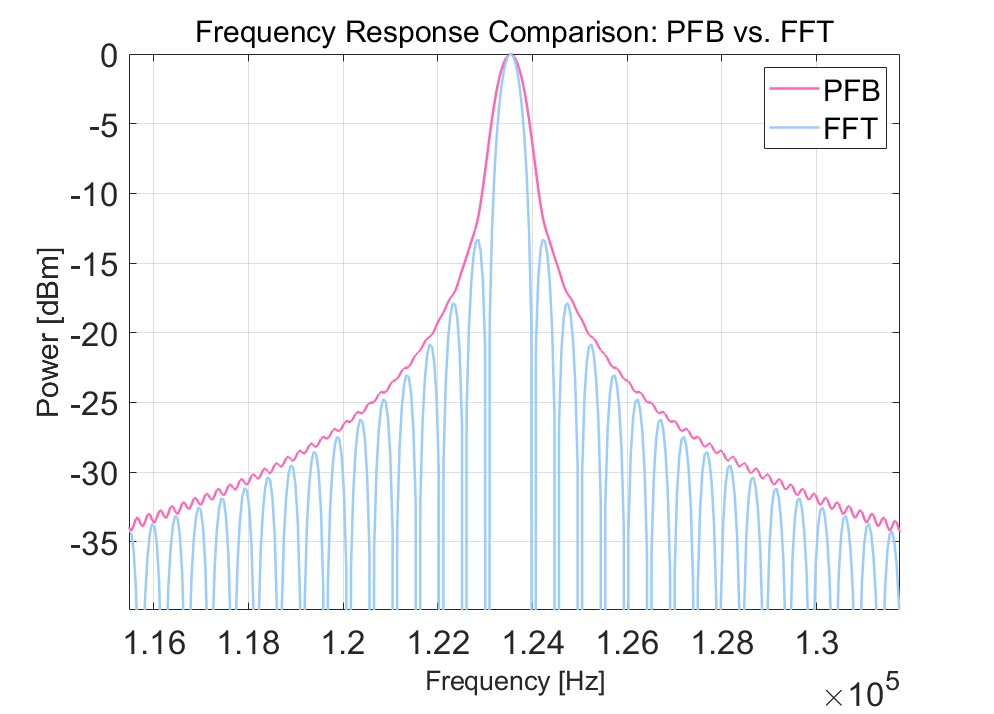}
\caption{Comparison of the single-bin frequency response between a Polyphase Filter Bank (PFB) and a direct FFT. The PFB shows significantly reduced spectral leakage and lower sidelobe levels compared to the FFT, resulting in improved channel separation due to the filtering effect of the polyphase window. A zoomed-in view highlights these differences clearly down to a noise floor of –50 dB.}
\label{figure:FFTvsPFB}
\end{figure}

The \(2^{19}\) input samples are initially processed through a size-\(2^{10}\) polyphase filter bank (PFB) for coarse channelization. Since MKID resonators are densely packed and may fall between standard FFT bin centers, the PFB architecture is critical. By integrating polyphase digital filters with the FFT, the PFB significantly reduces spectral leakage \cite{proakis2007digital} through improved channel selectivity and sidelobe suppression. This allows resonators that would otherwise be split across multiple bins in a basic FFT to be correctly resolved without information loss. A comparative plot of channel frequency responses for a standard FFT and a PFB, shown in Figure~\ref{figure:FFTvsPFB}, highlights this performance advantage.

Following coarse channelization, a size-\(2^9\) digital downconverter (DDC) provides fine channelization, isolating individual tones from the coarse bins with high resolution. The combined architecture achieves the required spectral precision while remaining compatible with FPGA resource constraints.

The coarse FFT size in channelization architectures is chosen to balance bandwidth coverage and the number of frequency channels, enabling initial spectral partitioning into manageable sub-bands. Subsequent fine FFT or DDC stages enhance frequency resolution within each coarse channel for precise tone recovery. While alternative FFT sizes are possible, typical selections represent a practical compromise optimizing frequency resolution, FPGA resource usage, and system complexity. Careful FFT size selection is essential for efficient signal processing in resource-constrained hardware, such as space-qualified FPGAs \cite{goodhart2013efficient, smith2022highly}.

\subsection{Critically Sampled and Weighted Overlap-Add PFBs}

Two principal PFB architectures are relevant for the coarse channelization stage in the MKID readout algorithm ~\cite{harris2022multirate}. A \textit{critically sampled} (CS) PFB maximizes efficiency by producing one output per input sample, resulting in contiguous, non-overlapping spectral coverage. This approach minimizes FPGA resource usage and is well suited for early development, prototyping, and structural verification under flight hardware constraints.

In contrast, the \textit{Weighted Overlap-Add} (WOLA) PFB introduces controlled overlap between adjacent channels, commonly by a factor of two, providing enhanced suppression of spectral leakage compared to CS PFBs~\cite{harris2022multirate}. This makes WOLA particularly effective for densely packed MKID arrays. However, it requires additional buffering, arithmetic resources, and phase correction logic. The trade-off is higher FPGA cost in exchange for improved spectral fidelity~\cite{harris2022multirate}.

\begin{figure}[h]
\centering
\includegraphics[width=8cm]{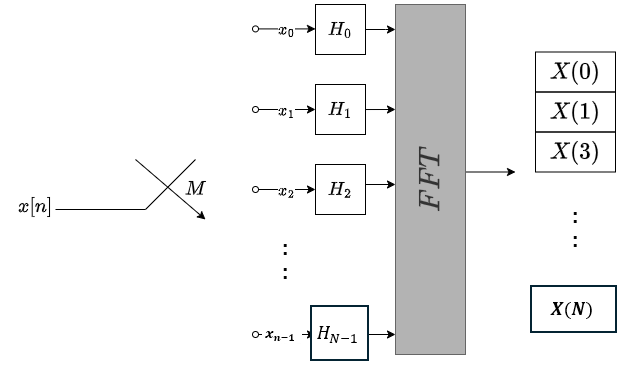}
\caption{Block diagram of the Critically-Sampled Polyphase Filter Bank (PFB) architecture. The input signal is segmented and filtered using polyphase components of a prototype filter, followed by an $N$-point FFT. This configuration provides contiguous, non-overlapping spectral bins with maximum efficiency and minimal FPGA resource usage.}
\label{figure:cs}
\end{figure}

\begin{figure}[h]
\centering
\includegraphics[width=8cm]{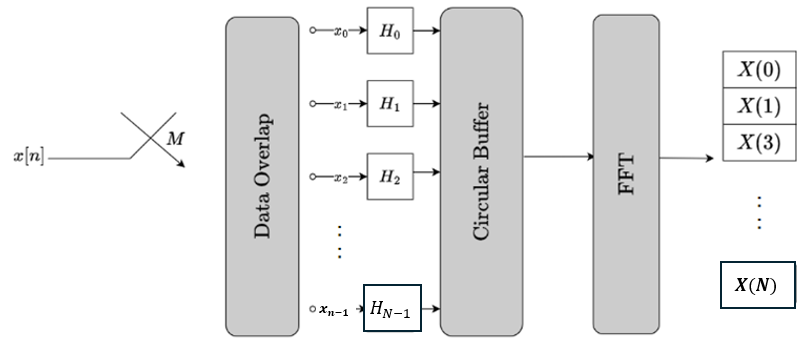}
\caption{Block diagram of the Weighted Overlap-Add (WOLA) Polyphase Filter Bank architecture. The input signal is processed in overlapping \(M/2\) segments before filtering with polyphase components. A circular buffer applies phase correction prior to streaming data into an \(N\)-point FFT. This overlap and weighting significantly reduce spectral leakage and enhance channel isolation, at the expense of increased FPGA resource usage.}
\label{figure:over}
\end{figure}

For the PRIMA and HWO missions, the development path begins with a critically sampled design to establish feasibility and validate the algorithm in hardware. The final implementation then transitions to a WOLA PFB to achieve the higher isolation and recovery precision required for dense MKID arrays.

\subsection{Architectural Differences Between CS and WOLA}

 The distinguishing characteristics of the two PFB architectures is seen in how input samples are grouped. In CS, groups of $M=N$ samples are processed directly, producing one output per input with no redundancy. In WOLA, the first $M/2$ samples are delayed and overlapped with the subsequent $M/2$, requiring buffering and reordering before the FFT stage. This induces a phase shift given by

\begin{equation}
\varphi(\omega_k) = \Delta t \cdot (-\omega_k),
\end{equation}

where $\Delta t$ is the delay and $\omega_k$ is the center frequency of bin $k$. For uniform sampling, this reduces to

\begin{equation}
\varphi(\omega_k) = \frac{n k}{M} 2\pi.
\end{equation}

In CS, where $n=M$, the phase shift is zero modulo $2\pi$. In WOLA, where $n=M/2$, odd-numbered channels experience a phase inversion that must be corrected through buffering. While this increases design complexity, the resulting improvement in channel isolation makes WOLA the architecture of choice for high-precision coarse channelization in PRIMA and HWO.

\medskip

In summary, the HWO staged development path begins with a critically sampled polyphase filter bank (PFB) to validate performance and resource efficiency, and advances to a weighted overlap-add (WOLA) PFB for flight deployment. This progression ensures both optimal FPGA resource utilization and the spectral precision required to meet the 10~kHz resolution target. This architecture delivers a space-qualified solution for high-density MKID array readout.

\section{FPGA-Based Design Methodology and Results}

The FPGA implementation of the polyphase filter bank (PFB) follows a disciplined three-step process adapted from heritage development for astrophysics missions. This structured methodology ensures performance efficiency, traceability, and compliance with spaceflight qualification requirements. The three stages consist of: (1) frequency planning, (2) fixed-point modeling and verification, and (3) HDL code generation and hardware implementation.

\subsection{Stage 1: Frequency Planning}

The first stage is the frequency planning phase, where the key design parameters are defined. These include the system sampling frequency $F_{s}$, the FFT length $N$ (which sets the number of channels), the decimation factor $M$ (typically $M=N$ for critically sampled designs and $M/2$ for WOLA), the desired stopband attenuation in dB, and the number of taps per channel. Collectively, these parameters establish the spectral resolution, oversampling ratio, and the total number of filter coefficients required for the prototype design. The requirements for PRIMA’s frequency plan are summarized in Table~\ref{table:PrototypeFIlter}. This disciplined approach ensures that the detector-level requirements are carried directly into the FPGA design.

\begin{table*}
\centering
\caption{Frequency Plan Parameters for Prototype Filter Design. \label{table:PrototypeFIlter}}
\renewcommand{\arraystretch}{1.33}
\begin{tabular}{l|c|c|l}
Parameter & CS & WOLA & Interpretation \\
\hline\hline
Sampling Frequency $F_{s}$ & 5 GHz & 5 GHz &  \\
FFT Length ($N$) & 1024 & 1024 & Channels \\
Downsampling ($M$) & M=1024 & $M/2$ = 512 & Filter Passband \\
Oversampling Ratio & 1 & 2 & $N/M$ \\
Taps & 4 & 4 &  \\
Number of Filter Coefficients & 4096 & 4096 & $Taps \times FFT$ \\
Stopband Attenuation & –60 dB & –60 dB &  \\
\hline
\end{tabular}
\medskip
\end{table*}

\subsection{Stage 2: Fixed-Point Modeling and Verification}

The second stage is the fixed-point modeling phase, implemented in MATLAB Simulink. Our current work is at this stage. A critically sampled polyphase filter bank (PFB) was implemented in Simulink using HDL-compatible blocks to support FPGA synthesis. The design employs a four-tap prototype low-pass filter, decomposed across 1024 phases, resulting in a 1024×4 coefficient matrix. Each row in this matrix corresponds to a distinct phase of the prototype filter, and each column represents a successive tap. This decomposition enables the filtering operation to be distributed across 1024 parallel branches, significantly reducing the computational load per channel. The outputs of these branches feed into a 1024-point FFT, forming the complete PFB architecture. 

\begin{figure}[h]
\centering
\includegraphics[width=9.5cm]{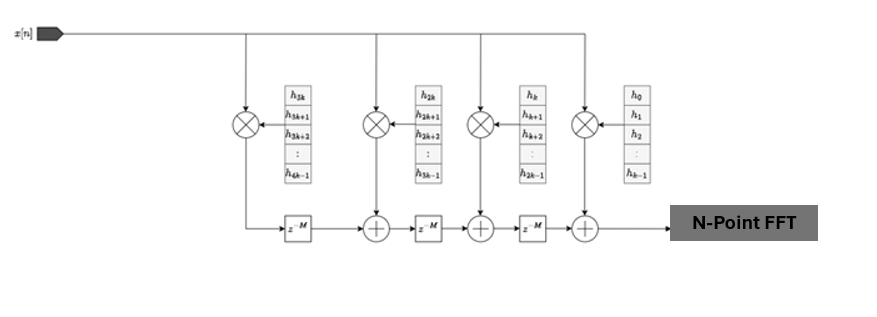}
\caption{Critically sampled PFB efficient hardware implementation: A single 4-tap subfilter operates on all filter coefficients stored in ROM arranged in a polyphase matrix, where each row corresponds to a distinct phase of the prototype filter. During each clock cycle, the subfilter processes one phase sequentially, enabling efficient real-time filtering with only four multiplications and additions per sample. This time-multiplexed approach significantly reduces resource utilization while maintaining full-rate throughput.}
\label{fig:fpga_cs}
\end{figure}

In Figure \ref{fig:fpga_cs}, to optimize resource usage in hardware, the PFB employs a time-multiplexed architecture in which all 1024 filter phases share a single subfilter. The prototype filter coefficients are arranged in a 1024×4 matrix, where each row corresponds to a distinct phase and each column represents one of the four filter taps. Rather than instantiating 1024 parallel filters, the design reuses a single 4×1 subfilter that is updated each clock cycle to process one row of the coefficient matrix at a time. This structure dramatically reduces the number of multipliers and adders required, enabling high channelization efficiency while maintaining the critically sampled configuration. Over 1024 clock cycles, the subfilter sequentially processes all phases of the input data stream, and the outputs are accumulated and passed into a 1024-point FFT. This serialized filtering approach achieves substantial hardware savings with minimal impact on throughput, making it highly suitable for resource-constrained FPGA environments.

\begin{figure}[h]
\centering
\includegraphics[width=9.5cm]{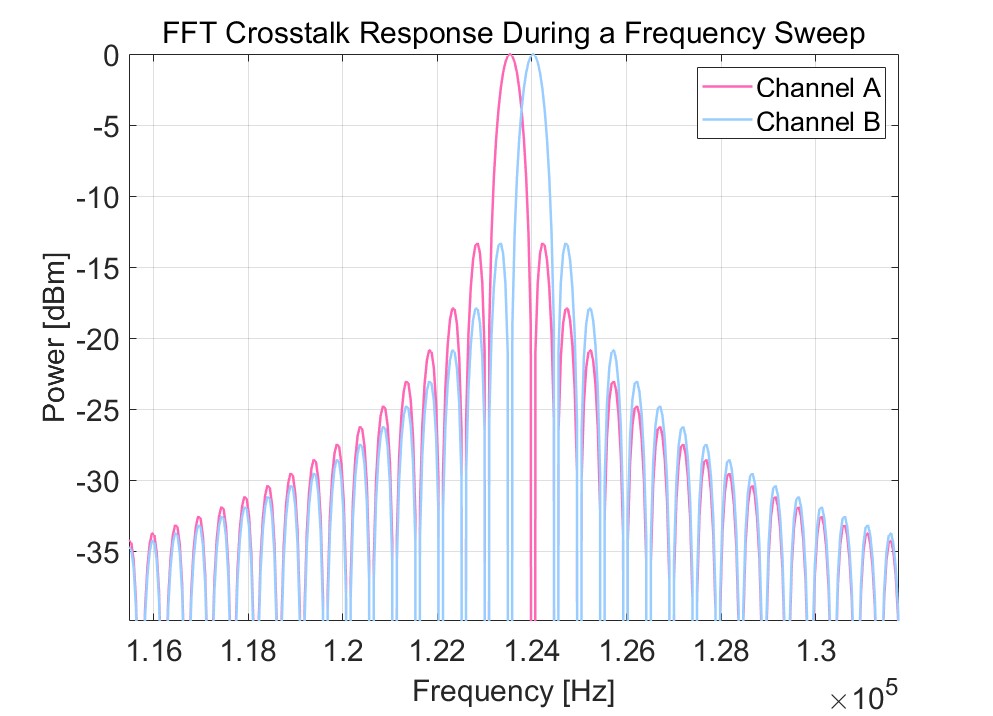}
\caption{FFT Crosstalk Response: The frequency response of the FFT-based channelizer during a continuous frequency sweep reveals significant sidelobe levels, indicative of spectral leakage and inter-channel crosstalk. These high sidelobes result from the rectangular windowing inherent in standard FFT processing, leading to poor spectral containment. These features are zoomed in to clearly illustrate the interchannel response, with the amplitude plotted down to a noise floor of $-50\,\mathrm{dB}$, allowing detailed comparison of adjacent channel interference and spectral containment.}
\label{figure:FFT Crosstalk}
\end{figure}

\begin{figure}[h]
\centering
\includegraphics[width=9.5cm]{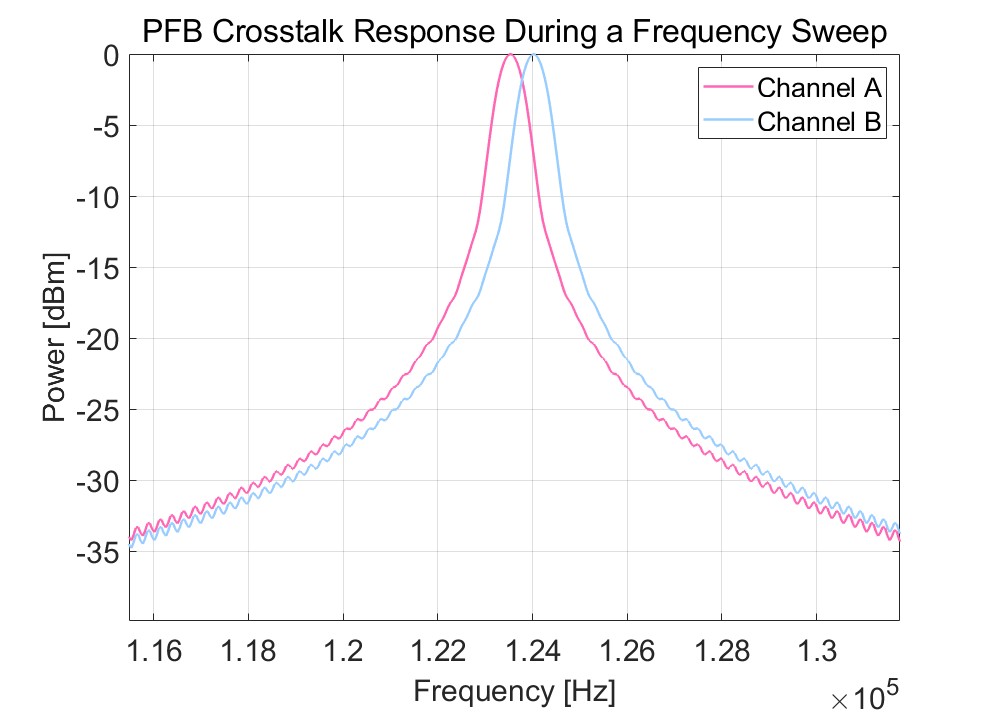}
\caption{PFB Crosstalk Response: The frequency response of the critically sampled PFB-based channelizer during a continuous frequency sweep reveals a significant reduction in sidelobe levels. This improvement is attributed to the polyphase filter, which applies windowed filtering to suppress sidelobes and minimize spectral leakage. These features are zoomed in to clearly illustrate the interchannel response, with the amplitude plotted down to a noise floor of $-50\,\mathrm{dB}$, allowing detailed comparison of adjacent channel interference and spectral containment.}
\label{figure:PFB_crosstalk}
\end{figure}

A full-band frequency sweep was performed on both the Simulink fixed-point CS PFB and the FFT-based channelizer to evaluate channel crosstalk and overall spectral performance. The resulting spectral responses highlight the superior frequency isolation and reduced inter-channel leakage achieved by the PFB (Figure \ref{figure:PFB_crosstalk}). In contrast, the FFT-based channelizer (Figure \ref{figure:FFT Crosstalk}) exhibits high sidelobes in its frequency response, an indication of poor spectral containment, leading to increased spectral leakage and reduced channel selectivity. These sidelobes are a result of the inherent rectangular windowing in FFT processing, which lacks the spectral tapering provided by the polyphase filtering in the PFB. Together, these results demonstrate the PFB’s effectiveness in delivering precise and reliable channelization, which is essential for high-density and high-fidelity signal processing applications.

\begin{figure}[h]
\centering
\includegraphics[width=0.53\textwidth,clip,trim=0in 0in 0in 0in]{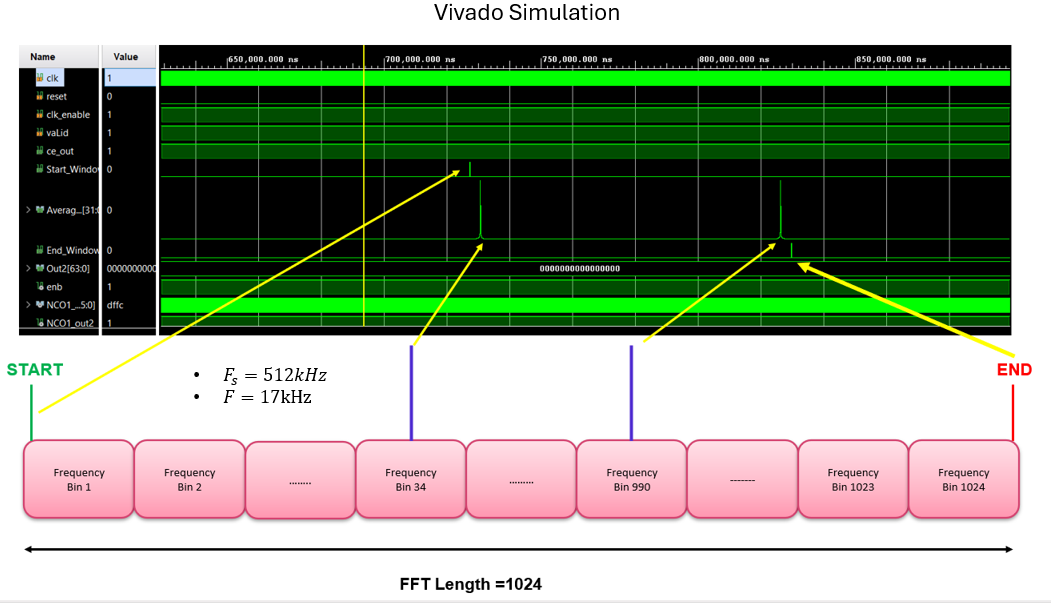}
\caption{Single-Tone Test Vivado Simulation: A sinusoidal test signal was generated using a Numerically Controlled Oscillator (NCO), implemented in hardware description language and embedded within the PFB system. The tone was synthesized with a sampling frequency \( F_s = 512\,\text{kHz} \) and a tone frequency \( F = 17\,\text{kHz} \), resulting in the signal aligning with FFT bin \( n = 34 \) of a 1024-point FFT.}
\label{figure:VivadoSimulation}
\end{figure}

\subsection{Stage 3: Hardware Implementation}

HDL code was generated from the fixed-point Simulink model and integrated within the AMD Vivado development environment, the critically sampled PFB architecture was implemented in hardware.  An embedded design combining the polyphase filterbank with an HDL numerically controlled oscillator (NCO) was developed and wrapped with an AXI interface to enable software control of frequency tuning words. This configuration allowed dynamic adjustment of the NCO frequency during testing. The Vivado Simulation results are displayed in Figure \ref{figure:VivadoSimulation} with the NCO configured with $Fs=512kHz$ and $F=17kHz$ according to the equation below which results in the FFT bin, $n=34$.

\begin{equation}
F=n\frac{F_{s}}{FFT Length}
\end{equation}

\begin{figure}[h!]
\centering
\includegraphics[width=0.48\textwidth,clip,trim=0in 0in 0in 0in]{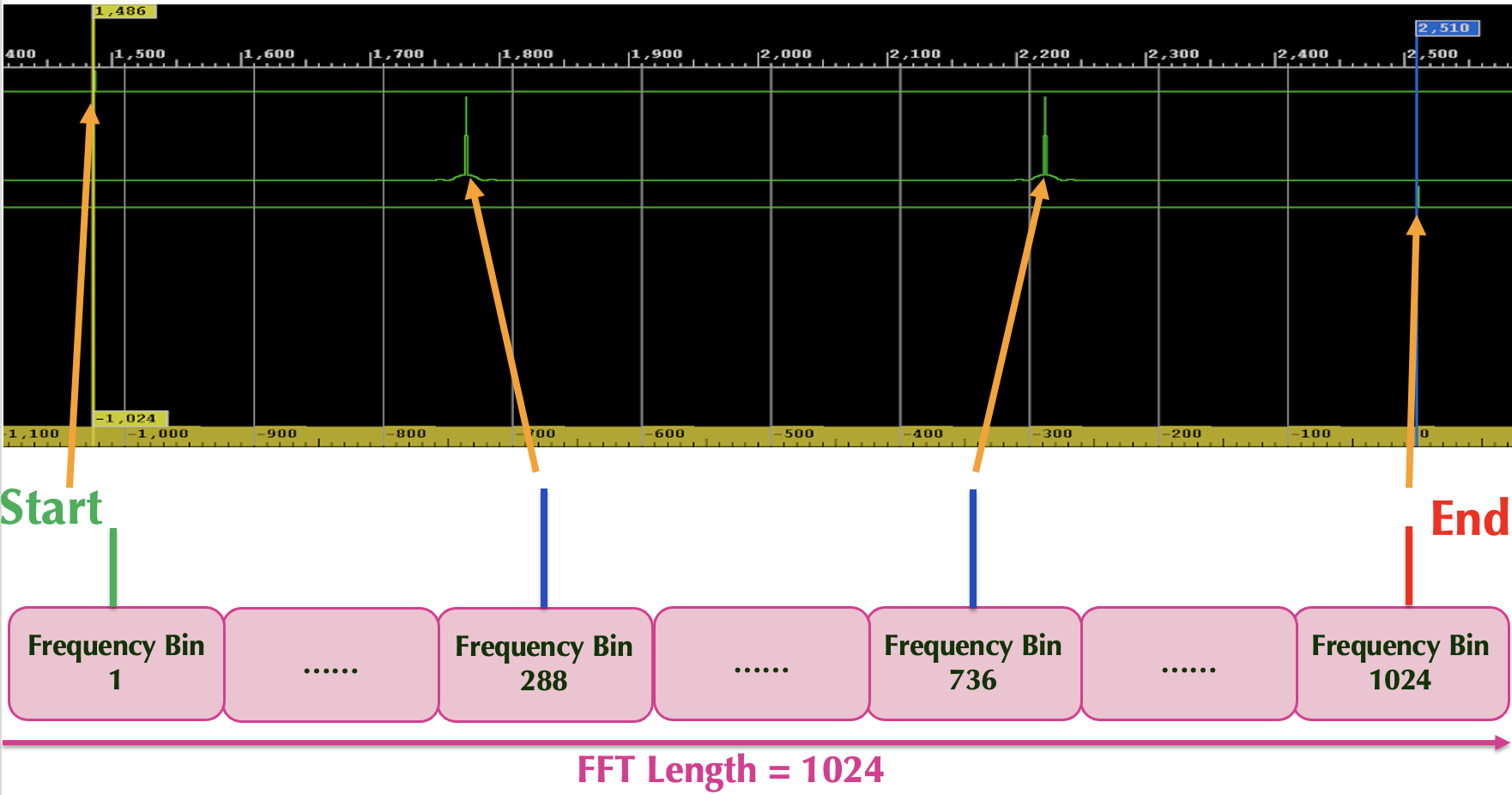}
\caption{Embedded Critically Sampled Polyphase Filterbank Hardware Implementation. 
The embedded PFB integrates a Numerically Controlled Oscillator (NCO), 
which is controlled via the AXI interface using Vitis software. 
The system operates with a sampling frequency of \( F_s = 128\,\mathrm{MHz} \) 
and an input tone frequency of \( F = 36\,\mathrm{MHz} \), corresponding to FFT bin \( n = 288 \) 
in a 1024-point FFT.}
\label{figure:emb_sys}
\end{figure}

Hardware deployment was conducted on both the Xilinx RFSoC and ZedBoard evaluation platforms. Signal data were captured using the Integrated Logic Analyzer (ILA) for real-time monitoring , shown in \ref{figure:emb_sys}. The captured results confirmed the correct operation of the channelizer and demonstrated its feasibility for real-time processing in resource-constrained FPGA environments typical of space-qualified applications.

The final step of this stage will be translation into hand-coded, space-qualified VHDL, with verification through bit-true test benches. This process ensures compliance with the stringent review, optimization, and reliability requirements of NASA flight hardware. Completion of this workflow advances the design along the Technology Readiness Level (TRL) pathway, providing a clear progression from algorithm development to space-qualified implementation.

\section{Future Work}

\begin{figure}[h!]
\centering
\includegraphics[width= 9 cm]{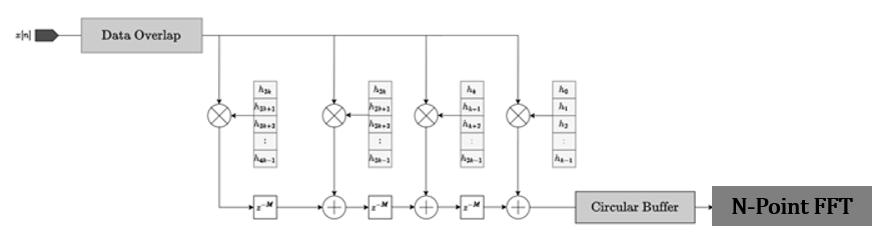}
\caption{Weighted Overlap and Add PFB Efficient Hardware Implementation: Hardware implementation of the weighted overlap-add (WOLA) architecture. Incoming samples are arranged with an overlap of \(M/2\) (half the downsampled length) before filtering. Filter coefficients are stored in a polyphase structure in ROM. Following filtering, phase correction is applied via a circular buffer prior to FFT processing. This efficient structure minimizes resource use while maintaining full-rate throughput.}
\label{figure:pfb_over}
\end{figure}

Future development will extend the critically sampled PFB to a weighted overlap-add (WOLA) architecture, as illustrated in Figure~\ref{figure:pfb_over}. To accommodate the 5~GHz input sampling rate, the PFB will be channelized into \(k\) parallel streams, each operating at a reduced clock rate of \(F_s / k\), facilitating timing closure within FPGA constraints. Spaceflight-grade VHDL will be developed from the Simulink model, with a focus on efficient utilization of DSP slices, optimized coefficient storage in block RAM, and the creation of in-flight reconfigurable software interfaces. 

Future development also includes selecting prototype windows and tap/overlap factors that meet the stated stopband and ripple targets; quantifying spectral behavior (3 dB bandwidth, passband ripple, sidelobe/leakage, and scalloping loss versus fractional-bin offset); defining instrument-appropriate minimum tone spacing from resonator linewidth and drift models and verifying adjacent-channel isolation and two-tone intermodulation to that specification; establishing a readout-noise budget below the front-end amplifier floor using converter quantization, clock-jitter, and phase-noise models; finalizing fixed-point word-length, rounding, and saturation allocations with an end-to-end SNR/SFDR budget and bit-true equivalence to the floating-point reference; and confirming throughput, latency, buffering, crest-factor management, and telemetry margins (e.g., SpaceWire framing/rates), together with FPGA utilization, timing closure, and power, for both CS and WOLA variants to demonstrate scalability to the target resonator count and aggregate bandwidth for space-based imaging and spectroscopy.

Building on the PRIMA baseline and extending to HWO requirements, this staged development path ensures that the polyphase filter bank architecture remains both scientifically capable and resource-efficient, ultimately delivering a space-qualified solution for high-density MKID readout in next-generation observatories.

\medskip

\bibliographystyle{IEEEtran}
\bibliography{report}

\end{document}